\documentclass[conference]{IEEEtran}
\usepackage{cite}

\ifCLASSINFOpdf
   \usepackage[pdftex]{graphicx}
\else
   \usepackage[dvips]{graphicx}
\fi
\usepackage{subfigure}

\usepackage{amssymb}
\usepackage[cmex10]{amsmath}
\usepackage{array}
\newtheorem{definition}{Definition}
\usepackage{centernot}
\usepackage{mathtools}
\DeclarePairedDelimiter{\floor}{\lfloor}{\rfloor}
\let\emptyset\varnothing

\begin{document}
%
\title{A Theoretical Framework for Understanding Mutation-Based Testing Methods}

\author{\IEEEauthorblockN{Donghwan Shin}
\IEEEauthorblockA{School of Computing\\
KAIST\\
Daejeon, Republic of Korea\\
Email: donghwan@se.kaist.ac.kr}
\and
\IEEEauthorblockN{Doo-Hwan Bae}
\IEEEauthorblockA{School of Computing\\
KAIST\\
Daejeon, Republic of Korea\\
Email: bae@se.kaist.ac.kr}}



\maketitle

\begin{abstract}
In the field of mutation analysis, mutation is the systematic generation of mutated programs (i.e., mutants) from an original program. The concept of mutation has been widely applied to various testing problems, including test set selection, fault localization, and program repair. However, surprisingly little focus has been given to the theoretical foundation of mutation-based testing methods, making it difficult to understand, organize, and describe various mutation-based testing methods.

This paper aims to consider a theoretical framework for understanding mutation-based testing methods. While there is a solid testing framework for general testing, this is incongruent with mutation-based testing methods, because it focuses on the correctness of a program for a test, while the essence of mutation-based testing concerns the differences between programs (including mutants) for a test.

In this paper, we begin the construction of our framework by defining a novel testing factor, called a test differentiator, to transform the paradigm of testing from the notion of correctness to the notion of difference. We formally define behavioral differences of programs for a set of tests as a mathematical vector, called a d-vector. We explore the multi-dimensional space represented by d-vectors, and provide a graphical model for describing the space. Based on our framework and formalization, we interpret existing mutation-based fault localization methods and mutant set minimization as applications, and identify novel implications for future work.
\end{abstract}


%
\IEEEpeerreviewmaketitle

\section{Introduction}\label{sec:intro}

In the field of mutation analysis, mutation is the systematic generation of mutated programs (i.e., mutants) from an original program. DeMillo et al. \cite{demillo1978hints} first proposed the notion of mutation for measuring the quality of a set of tests using mutants in the late 1970s. Mutation-based testing has been widely studied with the aim of addressing various testing problems, such as test set selection \cite{andrews2006using,li2009experimental,just2014mutants}, robustness testing \cite{schulte2014software}, fault localization \cite{papadakis2013metallaxis,moon2014ask}, and program repair \cite{debroy2010using,le2012genprog,debroy2014combining}. However, surprisingly little focus has been given to the theoretical foundation of mutation-based testing methods, making it difficult to understand, organize, and describe various mutation-based testing methods.

This paper aims to consider a theoretical framework for understanding mutation-based testing methods. A theoretical framework is a well-formed model of the general entities that are under investigation, which facilitates a clear understanding of the fundamentals of complex problems. For example, in the early 1980s Gourlay \cite{gourlay1983mathematical} organized the existing foundational studies in software testing\cite{goodenough1975toward,howden1976reliability,weyuker1980theories}, and defined a formal model for testing, called a testing system. This includes fundamental testing factors (i.e., programs, specifications, tests) and their formal relationships. Staats et al. \cite{staats2011programs} recently revisited the testing system by introducing a fourth testing factor, a test oracle, which had previously been implicitly considered. This testing system delivers a holistic view of the testing factors, and serves as a guide for discussions in studies such as \cite{staats2012understanding,polikarpova2013good,fraser2014achieving,barr2015oracle}. However, this is incongruent with mutation-based testing methods, because it focuses on the correctness of a program for a test, while the essence of mutation-based testing concerns the differences between programs (including mutants) for a test. 

For example, consider the recent mutation-based fault localization studies of Papadakis and Traon \cite{papadakis2013metallaxis} and Seokhyeon et al. \cite{moon2014ask}. In these studies, in order to find the exact locations of faults in a program (i.e., the original program with regard to mutants) it is essential to analyze the behavioral differences between mutants, the original program, and the specification (i.e., intended behavior), for a given set of tests. However, none of the existing theoretical frameworks can consistently describe these various behavioral differences. 

In this paper, we provide a solid theoretical framework for the notion of behavioral differences between programs in mutation-based testing. We begin the construction of our framework by defining a novel testing factor, called a test differentiator, to transform the paradigm of testing from the notion of correctness to the notion of difference. Using the test differentiator, we formally define behavioral differences between programs for a set of tests as a mathematical vector, called a d-vector. Based on the fact that a vector can be regarded as representing a point in a multidimensional space, we define the space of program behaviors for a set of tests, and explore the theoretical properties of that space. We conclude our theoretical framework with a graphical model to describe the space of behavioral differences. Based on our framework and formalization, we interpret existing mutation-based fault localization methods and mutant set minimization as applications, and identify novel implications for future work.

The remainder of this paper is structured as follows. Section \ref{sec:mutationBasedTesting} introduces some background material, and describes the scope of the mutation-based testing to be considered in this paper. Section \ref{sec:framework} describes our theoretical framework for mutation-based testing, including the new testing factor, the formal definition of behavioral differences, and our graphical model for behavioral differences. Section \ref{sec:applications} presents applications of the proposed formal framework. Section \ref{sec:conclusion} concludes the paper.

\section{Mutation-Based Testing}\label{sec:mutationBasedTesting}

Mutation analysis is a method for measuring the quality of the set of tests, using artificially injected faults (mutants) generated from the original program using predefined rules (mutation operators). If a mutant and the original program return different results for a test, then the test \textit{kills} the mutant. \textit{Mutation adequacy} is satisfied when all of the generated mutants are killed by the set of tests.

On the other hand, mutation testing provides a process for detecting faults in the original program. It starts with the generation of mutants from the original program, and tests are generated with the aid of automatic test generation methods. The generation of new tests and execution of live mutants are automatically repeated until the set of tests kills all of the mutants. This loop is the key element of mutation testing, which provides a set of tests satisfying the mutation adequacy condition. After all of the mutants are killed, the loop terminates and the original program is executed using the resulting test set to detect faults. 

There are two main hypotheses justifying mutation testing: The Competent Programmer Hypothesis (CPH) \cite{demillo1978hints} claims that the original program is made by competent programmers so the program has few simple faults. The coupling effect hypothesis \cite{offutt1992investigations} states that complex faults are coupled to simple faults so the tests that kill simple mutants will detect a large percentage of complex faults.

Several authors explored the theory of mutation in terms of test set selection for demonstrating the correctness of a program. Budd et al. \cite{budd1980theoretical} and Budd and Angluin \cite{budd1982two} presented the theoretical discussion on the test set selection problem in mutation testing. Morell \cite{morell1990theory} also discussed a theory of fault-based testing and considered the absence of prescribed faults in a program. On the other hand, many experimental studies have reported that tests satisfying the mutation adequacy are effective at detecting faults \cite{andrews2006using,li2009experimental,harman2011strong}. While these theoretical and empirical studies provide solid background for mutation testing in terms of the correctness of a program, to the best of our knowledge, there is no theoretical frameworks for mutation-based testing based on the differences of programs. 

By mutation-based testing we mean all mutation-based methods which attempt to solve various testing problems, not limited to fault detection problems. Mutation-based testing includes mutation-based fault detection, mutation-based fault localization, and mutation-based fault removal. In other words, if a testing method utilizes many mutants generated from an original program, then the method is regarded as one of the mutation-based methods.

Mutation-based testing differs from mutation analysis or mutation testing in terms of the way it utilizes mutants. Both mutation analysis and mutation testing are based on mutation adequacy, which means that they are based on the differences between each of mutants and an original program. However, mutation-based testing considers the differences not only between each of mutants and an original program, but also between each of mutants and a correct program (i.e., specification). The differences between a mutant and a correct program are particularly useful for fault localization and program repair \cite{le2012genprog,moon2014ask,debroy2014combining}. In this context, a correct program implies the intended behaviors of an original program for a given set of tests. While a correct program does not take the form of an executable program with source codes, in practice a human may play the role of the correct program, acting as human oracle. 

As Offutt noted in \cite{offutt2011mutation}, mutation applied to a program represents only one instance of a general class of applications. A general definition of mutation includes systematic changes to the syntax or to objects developed from the syntax. While this paper mainly focuses on program mutation, the application of results on this paper is not necessarily limited to programs, but can be extended to general syntax or objects. 
\section{Theoretical Framework}\label{sec:framework}

To serve as a guide for discussions for both theoretical and empirical studies, we focus on a concise framework rather than a theory including axioms, theorems, and proofs used to solve theoretical problems. In the following subsections, we explain the foundational concepts, definitions, and examples of our theoretical framework for understanding mutation-based testing methods. The applications of the theoretical framework are presented in Section \ref{sec:applications}.

\subsection{Basic Terms and Notations} %
Here, we will clarify the meanings of basic terms and notations used in this paper, including programs, specifications, tests, oracles, behaviors, and faults. We will then address the scope of mutation-based testing in comparison with mutation analysis and mutation testing.

In terms of programs, specifications, tests, and oracles, we will adopt the testing system used by Staats et al. \cite{staats2011programs}, because it has a solid historical background, and provides an intuitive portrayal of the general testing process. The testing system is a collection ($P$, $S$, $T$, $O$, $corr$, $corr_t$), where
\begin{itemize} 
\item $S$ is a set of specifications
\item $P$ is a set of programs
\item $T$ is a set of tests
\item $O$ is a set of oracles
\item $corr \subseteq P \times S$
\item $corr_t \subseteq T \times P \times S$
\end{itemize}
A specification $s\in S$ represents the true requirements of the \textit{program} $p\in P$. A \textit{test} $t\in T$ is a sequence of inputs accepted by some program. The correctness of a program is defined by the predicate $corr$. For a program $p\in P$ and a specification $s\in S$, the predicate $corr(p,s)$ implies that $p$ is correct with respect to $s$. Similarly, the predicate $corr_t$ implies correctness with respect to a test $t \in T$. In other words, $corr_t(t,p,s)$ holds if and only if $s$ holds for $p$ when running $t$. The values of $corr(p,s)$ and $corr_t(t,p,s)$ are theoretical, and are used to describe the relationship between testing and correctness. By definition, $corr_t(t,p,s)$ is always true for all tests if $corr(p,s)$ is true. Because the predicate $corr_t$ is theoretical, an \textit{oracle} $o\in O$ is defined as a predicate $o \subseteq T \times P$, which determines the passing or failure of $t$ for $p$ in practice. In general, it is assumed that $o$ approximates $corr_t$, even though it is not perfect.

In terms of the behaviors of programs in testing, we adopt the informal description given by Morell \cite{morell1990theory}: the behavior may include any test execution results, for example its output, its internal variables, its execution time, or its space consumption. For example, the correctness of $p$ for $t$ refers to the correctness of the behavior of $p$ for the execution of $t$. For the sake of simplicity, we assume that the behavior of a program for a test is deterministic and independent of the previous behavior. In the remainder of this paper, we will use the terms program and behavior of the program interchangeably. Note that we do not take a black-box perspective, because the behavior includes internal variables and more.

In terms of the faults of programs in testing, we say that $t$ detects a fault of $p$ when the behavior of $p$ for $t$ is inconsistent with $s$. In other words, a fault is a static defect in $p$ that results in an inconsistency between the intended behavior (i.e., $s$) and the behavior of $p$.

Throughout the remainder of this paper, $p_o\in P$ refers to the original program implemented to meet the specification $s$, and $p_s\in P$ refers to the projection of $s$ that represents the expected behaviors specified by $s$. While $p_s$ is not a real program, this is not a serious assumption, because we only require the behavior of $p_s$ for a given set of tests. In practice, a human may play the role of $s$ or $p_s$, acting as a human oracle. The notation $m\in M\subseteq P$ refers to a mutant generated from $p_o$, and the set of programs $M$ refers to a set of mutants. Note that $p_o$, $p_s$, and $m$ are general entities, and largely separated from any specifics such as programming languages or mutation methods.

\subsection{Test Differentiator: A New Testing Factor} %
The notion of difference is an abstract concept. In order to concretize and formalize the notion of difference in our framework, we define a new testing factor, called a test differentiator, as follows:
\vspace*{+.5\baselineskip}
\begin{definition}
A \textit{test differentiator} $d: T\times P\times P \to \{0,1\}$ is a function,\footnote{This function-style definition is replaceable by a predicate-style definition, such as $d\subseteq T\times P\times P$.} such that
\begin{displaymath}
d(t,p_x,p_y) = 
\begin{cases}
1\text{ }(true), & \text{if $p_x$ is \textit{different} with $p_y$ for $t$} \\ 
0\text{ }(false), & \text{otherwise}
\end{cases}
\end{displaymath}
for all tests $t\in T$ and programs $p_x,p_y\in P$.
\end{definition}
\vspace*{+.5\baselineskip}

By definition, a test differentiator concisely represents whether the behaviors of $p_x\in P$ and $p_y\in P$ are different for $t$. Note that we make no attempt to incorporate any specific definition of program differences in our framework. The specific definition of differences can only be decided in context. For example, while 0.3333 is different with 1/3 in the strict sense, 0.3333 will be regarded as the same as 1/3 in some cases. To keep things general, we consider a set of test differentiators $D$ that includes all possible test differentiators.

Having a test differentiator $d$ as the fundamental testing factor makes it possible to formalize many important concepts in mutation-based testing methods. For example, the notion of mutation adequacy, which is the essence of mutation-based testing, can be clearly and concisely formalized as follows:
\begin{equation}
\label{eq:mutationAdequacy}
\forall m\in M, \exists t\in TS, d(t,p_o,m).
\end{equation}
In other words, all mutants are killed by at least one test in the test suite $TS\subseteq T$. By the formalization of mutation adequacy in (\ref{eq:mutationAdequacy}), it is shown that mutation adequacy is determined not only by $p_o$, $m$, and $t$, but also $d$. For example, there is a spectrum of mutation approaches from a strong mutation \cite{demillo1978hints} to a weak mutation \cite{woodward1988weak}, depending on which $d$ is used. In a strong mutation analysis, a test $t$ kills a mutant $m$ when the output of $m$ differs from the output of the original program $p_o$ for $t$. In a weak mutation analysis, $t$ kills $m$ when the internal states of $m$ and $p_o$ are different for $t$. As a result, (\ref{eq:mutationAdequacy}) implies that the holistic view of $p_o$, $m$, $t$, and $d$ should be carefully considered to meet a certain level of mutation adequacy.

Consider an oracle $o$ and a differentiator $d$, they are similar in terms of their role in testing; $o$ implies the correctness of a program for a test, and $d$ implies the differences between programs for a test. In fact, for all $o\in O$, there are proper $d\in D$ and $p_s\in P$ where
\begin{displaymath}
\forall t\in T, \forall p\in P, o(t,p) \Leftrightarrow \neg d(t,p,p_s)
\end{displaymath}
In other words, $d$ can play the role of $o$ with the aid of $p_s$. For example, the correctness of a program $p$ for a test $t$ is written by not only $o(p,t)$ but also $d(t,p,p_s)$. However, it is clear that $o$ cannot play the role of $d$ in general. This means that $d$ is more general than $o$. In the rest of this paper, $d$ is consistently used without $o$.

We should note that Staats et al. \cite{staats2011programs} formulated mutation adequacy as $\forall m\in M, \exists t\in TS, \neg o(t,m)$. However, as they already mentioned, their formulation is inaccurate, because general mutation adequacy does not include the term $o(t,m)$, which implies the correctness of $m$ for $t$. Mutation adequacy is based on the differences between $p_o$ and $m$ for $t$, which is exactly captured by $d(t,p_o,m)$.

\subsection{Behavioral Difference} %
Before we formally describe behavioral differences between programs for a set of tests, we introduce a \textit{test vector}, to formalize an ordered set of tests as follows:
\vspace*{+.5\baselineskip}
\begin{definition} 
A \textit{test vector} $\textbf{t}=\langle t_1, t_2, ..., t_n \rangle \in T^n$ is a vector where $t_i \in T$.
\end{definition}
\vspace*{+.5\baselineskip}

A test vector $\textbf{t} \in T^n$ is the same as a test suite $TS$ with size $n$, except that $\textbf{t}$ contains numbered tests. For example, two tests $t_x, t_y\in TS$ could form a test vector $\mathbf{t}=(t_1, t_2) \in T^2$. In this paper, bold letters represent vector forms. 

With the aid of $d$ and $\mathbf{t}$, we define \textit{d-vectors}, which formulates behavioral differences, as follows:
\vspace*{+.5\baselineskip}
\begin{definition}\label{def:d-vector} 
A \textit{d-vector} $\mathbf{d}:T^n\times P\times P \to \{0,1\}^n$ is an $n$-dimensional vector, such that
\begin{displaymath}
\mathbf{d}(\mathbf{t},p_x,p_y) = \langle d(t_1,p_x,p_y), ..., d(t_n,p_x,p_y) \rangle
\end{displaymath}
for all $\mathbf{t}\in T^n$, $d\in D$, and $p_x,p_y\in P$.
\end{definition}
\vspace*{+.5\baselineskip}

By definition, a d-vector $\mathbf{d}(\mathbf{t},p_x,p_y)$ represents the behavioral differences between $p_x$ and $p_y$ for all tests in $\mathbf{t}$ in vector form. In other words, $\mathbf{d}(\mathbf{t},p_x,p_y)$ effectively indicates the tests $t\in \mathbf{t}$ for which the two programs $p_x$ and $p_y$ exhibit different behaviors. For example, if $d(t_i,p_x,p_y)=1$ for some test $t_i \in \mathbf{t}$, this means that $p_x$ and $p_y$ are different for the particular test $t_i$ contained in the set of tests $\mathbf{t}$.

\begin{figure}
	\centering
	\includegraphics[width=0.95\linewidth]{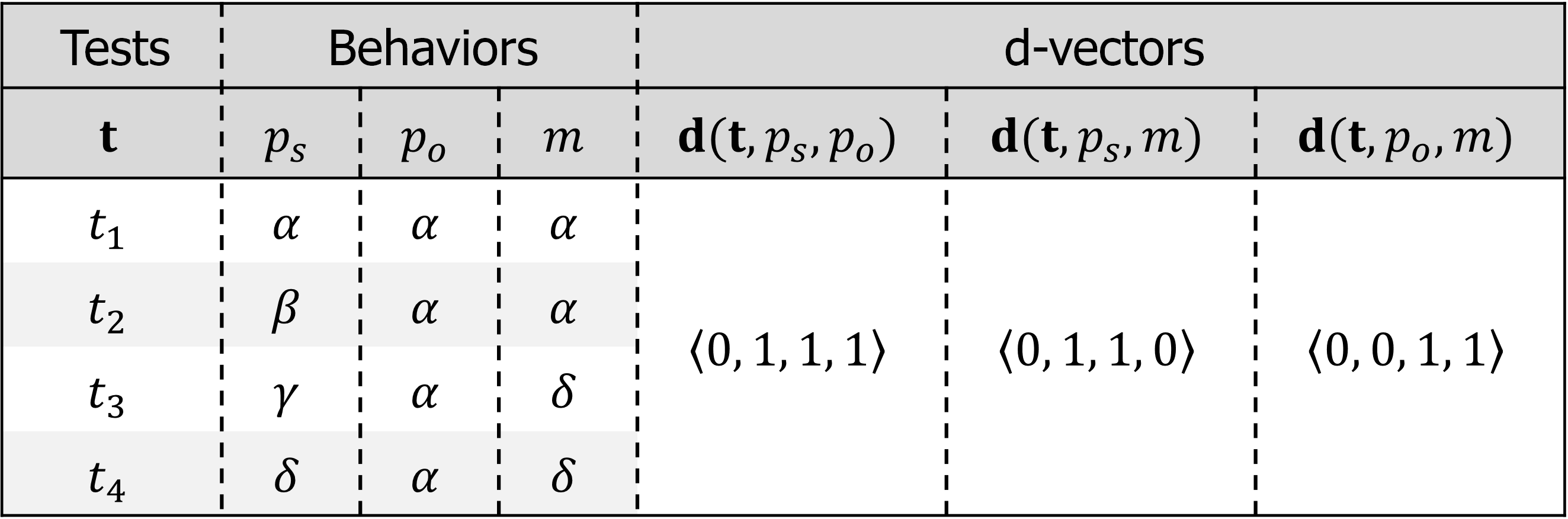}
	\caption{Running example for tests, program behaviors, and d-vectors}
	\label{fig:ex1}
\end{figure}

In Figure \ref{fig:ex1}, in a running example that we shall refer to through the remainder of Section \ref{sec:framework}, we present the program behaviors of $p_s$, $p_o$, and $m$ for $\mathbf{t}=\langle t_1, t_2, t_3, t_4 \rangle$ with $d$. Each behavior of a program for a test is abstracted by a Greek letter, and $d$ determines a difference between behaviors by a difference between Greek letters. On the right-hand side, there are d-vectors that represent the behavioral differences among the programs for the tests.

In the example, $\mathbf{d}(\mathbf{t},p_s,p_o)$ is equal to $\langle 0,1,1,1 \rangle$, because $d(t_1,p_s,p_o)=0$, $d(t_2,p_s,p_o)=1$, $d(t_3,p_s,p_o)=1$, and $d(t_4,p_s,p_o)=1$, respectively. Note that all of the behavioral differences among $p_s$, $p_o$, and $m$ are represented by d-vectors. 

A d-vector provides a quantitative difference by taking a mathematical norm of vectors. In the example, for the d-vector $\mathbf{d}(\mathbf{t},p_s,p_o) = \langle 0,1,1,1 \rangle$, the Manhattan norm\footnote{In general, considering the behavior of a program for a test as a string, the quantitative difference between two behaviors is measurable by the Hamming distance between two strings for the behaviors. Interestingly, it is consistent with the Manhattan norm of a d-vector: the Hamming distance between the behaviors of $p_x$ and $p_y$ for $\mathbf{t}$ is equivalent to the Manhattan norm of the d-vector $\mathbf{d}(\mathbf{t},p_x,p_y)$.} gives the quantitative difference as $0+1+1+1=3$, in terms of $d$. This means that the behavioral difference between $p_s$ and $p_o$ is $3$, quantitatively, in terms of the given $\mathbf{t}$ and $d$. This is written as $||\mathbf{d}(\mathbf{t},p_s,p_o)|| = 3$.

\subsection{Position: A New Interpretation of Behavioral Differences}\label{sec:position}
A vector can be regarded as representing the position of a point in a multi-dimensional space. For example, considering a $n$-dimensional space, a vector $\mathbf{v}=\langle v_1,...,v_n \rangle$ represents the point whose position in the $i$-th dimension (relative to the origin of the space) is $v_i$, for $i=\{1,...,n\}$. In this way, we can think of a d-vector as the representation of a position in a multi-dimensional space. We introduce this new interpretation of d-vectors as follows:
\vspace*{+.5\baselineskip}
\begin{definition}\label{def:position} 
The \textit{position} of a program $p_x$ relative to another program $p_r$ in a multi-dimensional space corresponding to a set of tests $\mathbf{t}$ is
\begin{displaymath}
\mathbf{d}_{p_r}^{\mathbf{t}}(p_x) = \mathbf{d}(\mathbf{t},p_r,p_x),
\end{displaymath}
where $\mathbf{d}(\mathbf{t},p_r,p_x)$ is the d-vector between $p_x$ and $p_r$ for $\mathbf{t}$, with regard to $d$. This multi-dimensional space is called the \textit{program space} induced by $(\mathbf{t}, p_r, d)$, where $\mathbf{t}$ corresponds to the set of dimensions, $p_r$ corresponds to the origin, and $d$ corresponds to the notion of differences between positions.
\end{definition}
\vspace*{+.5\baselineskip}

In other words, for all $t \in \mathbf{t}$, the behavioral difference between $p_r$ and $p_x$ for $t$ is indicated by the position of a program $p_x$ relative to the origin $p_r$ in the dimension $t$. Because $d$ returns either $0$ or $1$, there are only two possible positions in each dimension: the same position as the origin (i.e,. 0) and a different position from the origin (i.e., 1). It means that the semantics of a program $p_x$ in the space of $(\mathbf{t},p_r,d)$ is indicated by the $n$-bit binary vector $\mathbf{d}_{p_r}^{\mathbf{t}}(p_x)$ where $n = |\mathbf{t}|$.

The origin of a program space is important, because it determines the meaning of positions in the program space. In other words, the origin determines the meaning of the program space. For example, if the correct program $p_s$ is used for the origin, then the position $\mathbf{d}_{p_s}^{\mathbf{t}}(p_x)$ indicates how correct the program $p_x$ is with regard to $\mathbf{t}$. On the other hand, if the original program $p_o$ is used for the origin, and a mutant $m$ generated from $p_o$ is used for the target program $p_x$, then the position $\mathbf{d}_{p_o}^{\mathbf{t}}(m)$ indicates the killing of $m$ with regard to $\mathbf{t}$.

Conceptually, the position of a program in an $n$-dimensional program space translates the program behavior as an $n$-bit binary string. Each bit represents the behavioral difference between the program and another program at the origin of the space. In our example, the position of $m$ relative to $p_o$ is $\mathbf{d}_{p_o}^{\mathbf{t}}(m) = \langle 0,0,1,1 \rangle$. This means that $m$ is represented by $0011$ in the program space $(\mathbf{t},p_o,d)$. Such a concise representation makes it favorable to consider positions rather than d-vectors.

It is worthwhile to consider the norm of a position as well. The norm of the position of $p_x$ relative to $p_r$ naturally indicates the distance from $p_r$ to $p_x$ in the program space. For example, the norm of the position of $m$ relative to $p_o$ is $||\mathbf{d}_{p_o}^{\mathbf{t}}(m)|| = 2$ which means that the distance from $p_o$ to $m$ is 2. For an arbitrary program $p$ including mutants, $||\mathbf{d}_{p_s}^{\mathbf{t}}(p)||$ indicates the incorrectness of $p$ with respect to $\mathbf{t}$ and $d$. For an arbitrary mutant $m$, $||\mathbf{d}_{p_o}^{\mathbf{t}}(m)||$ indicates the easiness of killing $m$ with respect to $\mathbf{t}$ and $d$.

In our running example, there are two d-vectors $\mathbf{d}(\mathbf{t},p_s,p_o)$ and $\mathbf{d}(\mathbf{t},p_s,m)$. These indicate the two positions $\mathbf{d}_{p_s}^{\mathbf{t}}(p_o)$ and $\mathbf{d}_{p_s}^{\mathbf{t}}(m)$ in the same program space, by Definition \ref{def:position}. In other words, we have a four-dimensional space, the origin is $p_s$, and the two programs $p_o$ and $m$ are in $\langle 0,1,1,1 \rangle$ and $\langle 0,1,1,0 \rangle$, respectively. $m$ is closer to the origin $p_s$ than $p_o$ which means that $m$ is more correct than $p_o$. This shows that mutation can generate \textit{partially} correct mutants. This idea is used to the foundational concept for mutation-based fault localization and mutation-based program repair \cite{le2012genprog,moon2014ask,debroy2014combining}. Considering positions of many mutants makes it easy to understand and discuss the mutation-based methods. We will shows the specific application in Section \ref{sec:MBFL}.

\subsection{Different Positions and Different Behaviors}\label{sec:differentPositions}
In this subsection, we discuss the relationship between the positional difference of two programs and their behavioral difference. Let us consider two arbitrary positions in the same program space, in terms of Definition \ref{def:position}. Because the position of a program indicates the behavioral difference of the program relative to the origin, it is expected that there is a relationship between programs' positions and behaviors. For example, if two programs are in different positions in one dimension, it implies that the two programs have different behaviors for the corresponding test. This fact can be generalized as follows:
\begin{equation}
\label{eq:pos2diff}
(\mathbf{d}_{p_r}^{\mathbf{t}}(p_x) \neq \mathbf{d}_{p_r}^{\mathbf{t}}(p_y)) \implies (\mathbf{d}(\mathbf{t},p_x,p_y) \neq \mathbf{0}),
\end{equation}
for all $p_x,p_y,p_r\in P, d\in D,$ and $\mathbf{t}\in T^n$, with arbitrary $n$. In (\ref{eq:pos2diff}), the left-hand side (LHS) implies that $p_x$ and $p_y$ are in different positions in the program space of $(\mathbf{t},p_r,d)$. The right-hand side (RHS) implies that $p_x$ and $p_y$ have different behaviors for $\mathbf{t}$. Roughly speaking, (\ref{eq:pos2diff}) implies that the difference of positions of programs guarantees the difference of behaviors for the programs. The proof of this is omitted, because it is trivial. By (\ref{eq:pos2diff}), it is safe to conclude that programs in different positions have different behaviors for a given set of tests in a space. 

Note that the inverse of (\ref{eq:pos2diff}) does not hold. In other words, even if two programs are in the same position in a program space, this does not imply that the two programs have the same behaviors for the tests corresponding to the program space. This can be formalized as follows:
\begin{equation}
\label{eq:diff2pos}
(\mathbf{d}_{p_r}^{\mathbf{t}}(p_x) = \mathbf{d}_{p_r}^{\mathbf{t}}(p_y)) \centernot\implies (\mathbf{d}(\mathbf{t},p_x,p_y)=\mathbf{0}).
\end{equation}
Again, this holds for all $p_x,p_y,p_r\in P, d\in D,$ and $\mathbf{t}\in T^n$, with arbitrary $n$. In (\ref{eq:diff2pos}), the reason why the LHS does not imply the RHS is because of the case where $p_x$, $p_y$, and $p_r$ are all different to each other. An example of this is presented in Figure \ref{fig:ex1}. Consider $\mathbf{t'} = \langle t_3 \rangle$, where the three programs $p_s$, $p_o$, and $m$ have different behaviors. However, the position of $p_o$ relative to $p_s$ is the same as the position of $m$ relative to $p_s$, because $\mathbf{d}_{p_s}^{\mathbf{t'}}(p_o) = \mathbf{d}_{p_s}^{\mathbf{t'}}(m) = \langle 1 \rangle$.

To summarize, the position of a program in a program space indicates its behavioral difference with respect to the origin of the program space. The meaning of the position depends on what program is used as the origin of the space. The meaning of the difference between two positions is related with the behavioral differences of programs in those positions. If two programs are in different positions, this guarantees that the two programs' behaviors are different. However, if the positions are not different, this does not guarantee that the two programs' behaviors are equal.


\subsection{Formal Relation on Positions}\label{sec:relation}
In the previous section, we discussed the relationship between the difference of positions and the difference of behaviors. In this section, we introduce the formal relation on positions, called the \textit{deviance} relation, as follows:

\vspace*{+.5\baselineskip}
\begin{definition} 
\label{def:deviance}
For a program space defined by $(\mathbf{t}, p_r, d)$, the position $\mathbf{d}_{p_r}^{\mathbf{t}}(p_y)$ is \textit{transitively deviant from} the position $\mathbf{d}_{p_r}^{\mathbf{t}}(p_x)$ by $\mathbf{t_d}\subseteq \mathbf{t}$, if the following conditions hold:

(1) $\forall t\in \mathbf{t}-\mathbf{t_d}, d(t,p_r,p_x) = d(t,p_r,p_y)$,

(2) $\forall t\in \mathbf{t_d}, d(t,p_r,p_x)=0$,

(3) $\forall t\in \mathbf{t_d}, d(t,p_r,p_y)=1$, \\
for all $p_x,p_y\in P$. This is written as $\mathbf{d}_{p_r}^{\mathbf{t}}(p_x) \xrightarrow{\mathbf{t_d}} \mathbf{d}_{p_r}^{\mathbf{t}}(p_y)$ or simply $\mathbf{p_x} \xrightarrow{\mathbf{t_d}} \mathbf{p_y}$. 
\end{definition}
\vspace*{+.5\baselineskip}

In other words, (1) the position of $p_x$ is the same as the position of $p_y$ in all dimensions except for $\mathbf{t_d}$, (2) the position of $p_x$ in $\mathbf{t_d}$ dimensions is the same as the origin, (3) the position of $p_y$ in $\mathbf{t_d}$ dimensions is the opposite of the origin. Here, $\mathbf{t_d}$ represents every test $t\in \mathbf{t}$ where $d(t,p_x,p_y)=1$. When the triple $(\mathbf{t}, p_r, d)$ is not the main concern, we shorten the notation for the position of a program $p$ to the position (vector) $\mathbf{p}$.

The deviance relation indicates how tests influence positions in a testing process. In general, a set of tests increases in size to become more effective at detecting faults. This growth of the test set is described by $\mathbf{t_d}$ in Definition \ref{def:deviance}: $\mathbf{t_d}$ makes $\mathbf{p_y}$ deviant from $\mathbf{p_x}$. For example, if a correct program $p_s$ is given by $\mathbf{p_x}$, then $\mathbf{t_d}$ becomes the set of tests that detects faults in every program in $\mathbf{p_y}$. On the other hands, if an original program $p_o$ is at $\mathbf{p_x}$, then $\mathbf{t_d}$ becomes the set of tests that kills every mutant in $\mathbf{p_y}$.

The deviance relation is asymmetric. For all positions $\mathbf{p_x},\mathbf{p_y}$ and tests $\mathbf{t_d}$, the following is true:
\begin{displaymath}
\mathbf{p_x} \xrightarrow{\mathbf{t_d}} \mathbf{p_y} 
\implies \neg ( \mathbf{p_y} \xrightarrow{\mathbf{t_d}} \mathbf{p_x} )
\end{displaymath}

Interestingly, it is not only asymmetric but also transitive. For all positions $\mathbf{p_x},\mathbf{p_y},\mathbf{p_z}$ and tests $\mathbf{t_x},\mathbf{t_y}$, the following is true:
\begin{displaymath}
(\mathbf{p_x} \xrightarrow{\mathbf{t_x}} \mathbf{p_y}) \wedge (\mathbf{p_y} \xrightarrow{\mathbf{t_y}} \mathbf{p_z}) \implies (\mathbf{p_x} \xrightarrow{\mathbf{t_x} \text{ or } \mathbf{t_y}} \mathbf{p_z}).
\end{displaymath}
In other words, all positions deviant from $\mathbf{p_y}$ are also deviant from $\mathbf{p_x}$ if $\mathbf{p_y}$ is deviant from $\mathbf{p_x}$. This transitivity of the deviance relation on positions is closely related to the redundancy of mutants. For example, it may be the case that 
\begin{displaymath}
\mathbf{p_0} \xrightarrow{t_1} \mathbf{m_1} \xrightarrow{t_2} \mathbf{m_2} \xrightarrow{t_3} \mathbf{m_3} \rightarrow \cdots \xrightarrow{t_n} \mathbf{m_n},
\end{displaymath}
for an original program $p_o$, mutants $m_1,m_2,\cdots,m_n$, and tests $t_1,t_2,\cdots,t_n$. In this case, $t_1$ is enough to make all positions deviant from $\mathbf{p_0}$ because of the transitivity of the deviance relation, and this leads that $t_1$ is enough to kill all mutants. In other words, $m_2,\cdots, m_n$ are redundant by $m_1$ with respect to the given tests in this case. Because there is a well-defined method to find redundant mutants, introduced by Ammann et al. \cite{ammann2014establishing}, we analyze this using our framework in Section \ref{sec:DMSG}.

\subsection{Position Deviance Lattice}\label{sec:graph}
The deviance relation naturally forms a lattice of positions\footnote{Strictly, the deviance relation provides a strict partial order which is not a lattice but a directed acyclic graph. However, it is easy to consider the corresponding non-strict partial order given by the ``deviant from or equal to" relation. This non-strict partial order is a lattice, and we simply saying that the lattice is formed by the deviance relation on positions.}. We introduce the \textit{Position Deviance Lattice} (PDL) which shows the positions and their deviance relationships as follows:
\vspace*{+.5\baselineskip}
\begin{definition}\label{def:graph} 
For a program space given by $(\mathbf{t}, p_r, d)$, a \textit{position deviance lattice} is a directed graph $G=(N,E)$ consists of

(1) $N = \{ \mathbf{p}$ $|$ $\mathbf{p}=\mathbf{d}_{p_r}^{\mathbf{t}}(p)$ for $p\in PS \}$,

(2) $E = \{(\mathbf{p_x},\mathbf{p_y})$ $|$ $\mathbf{p_x} \xrightarrow{t} \mathbf{p_y}$ for single $t\in \mathbf{t} \}$ \\ 
where $(\mathbf{p_x},\mathbf{p_y})\in E$ refers to the directed edge from $\mathbf{p_x}$ to $\mathbf{p_y}$.
\end{definition}
\vspace*{+.5\baselineskip}

In other words, (1) the set of nodes includes all possible positions in the program space corresponding to $(\mathbf{t}, p_r, d)$, (2) the set of edges includes every directed pair of positions having the deviance relation with a single test. PDL represents the frame of the $n$-dimensional program space by its positions and deviance relations. Note that an $n$-dimensional program space contains $2^n$ possible positions, because each dimension contributes two positions. 

\begin{figure}
	\centering
	\includegraphics[width=0.9\linewidth]{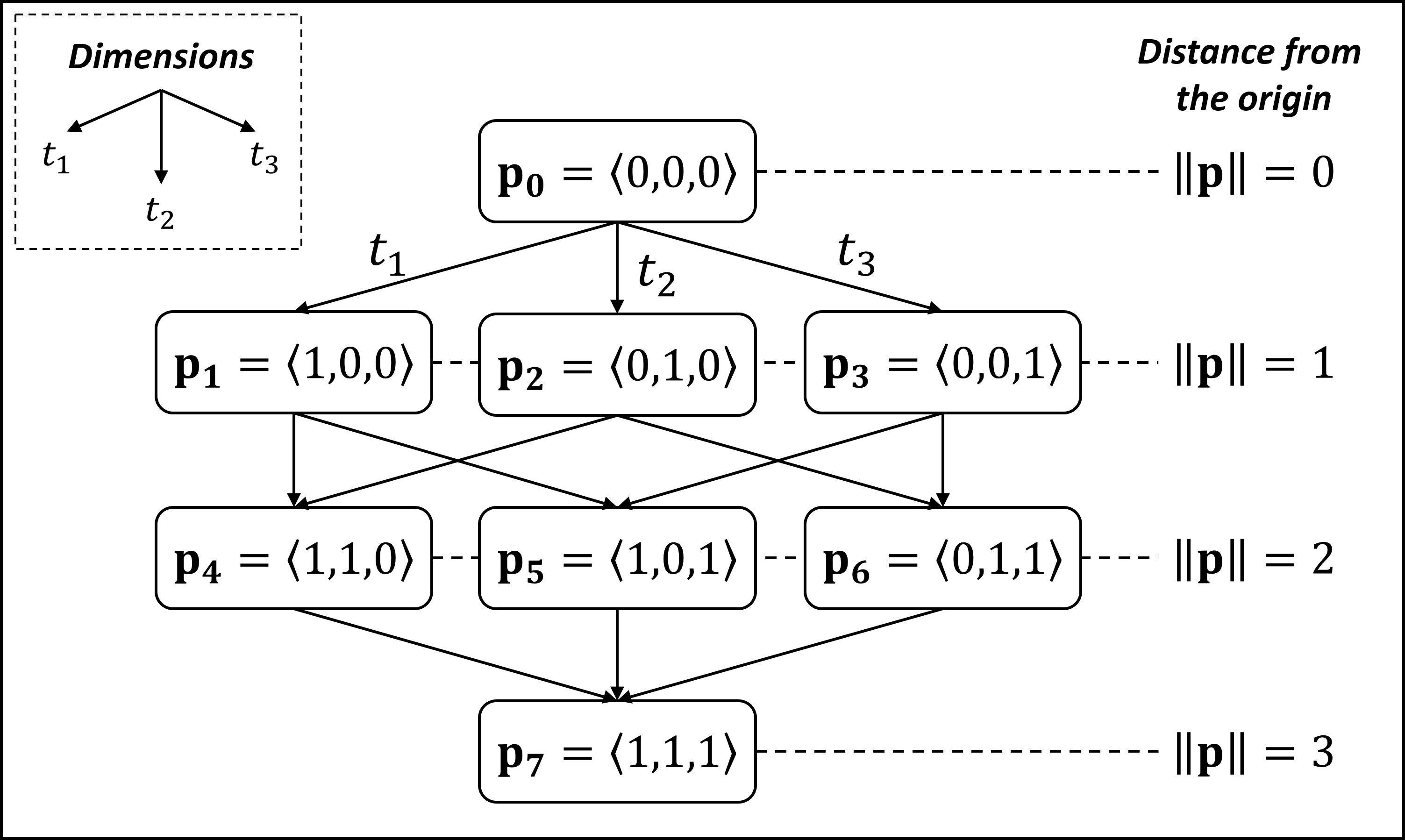}
	\caption{Position deviance graph for a three-dimensional program space}
	\label{fig:graph}
\end{figure}

For example, Figure \ref{fig:graph} presents the PDL that illustrates the frame of a three-dimensional program space containing eight positions. The arrows in the leftmost dashed-line box indicate the directions of the three dimensions $\mathbf{t}=\langle t_1,t_2,t_3 \rangle$. The main graph illustrates all of the possible positions $\mathbf{p_0}, \cdots, \mathbf{p_7}$ in the program space and their deviance relations. The values in the rightmost side indicate positions' distance from the origin.

PDL is useful for analyzing mutation-based methods. For example, we can analyze and improve the theoretical foundation for the mutant set minimization recently established by Ammann et al. \cite{ammann2014establishing}. We will explain this process in Section \ref{sec:DMSG}.

\begin{figure}
	\centering
	\includegraphics[width=0.95\linewidth]{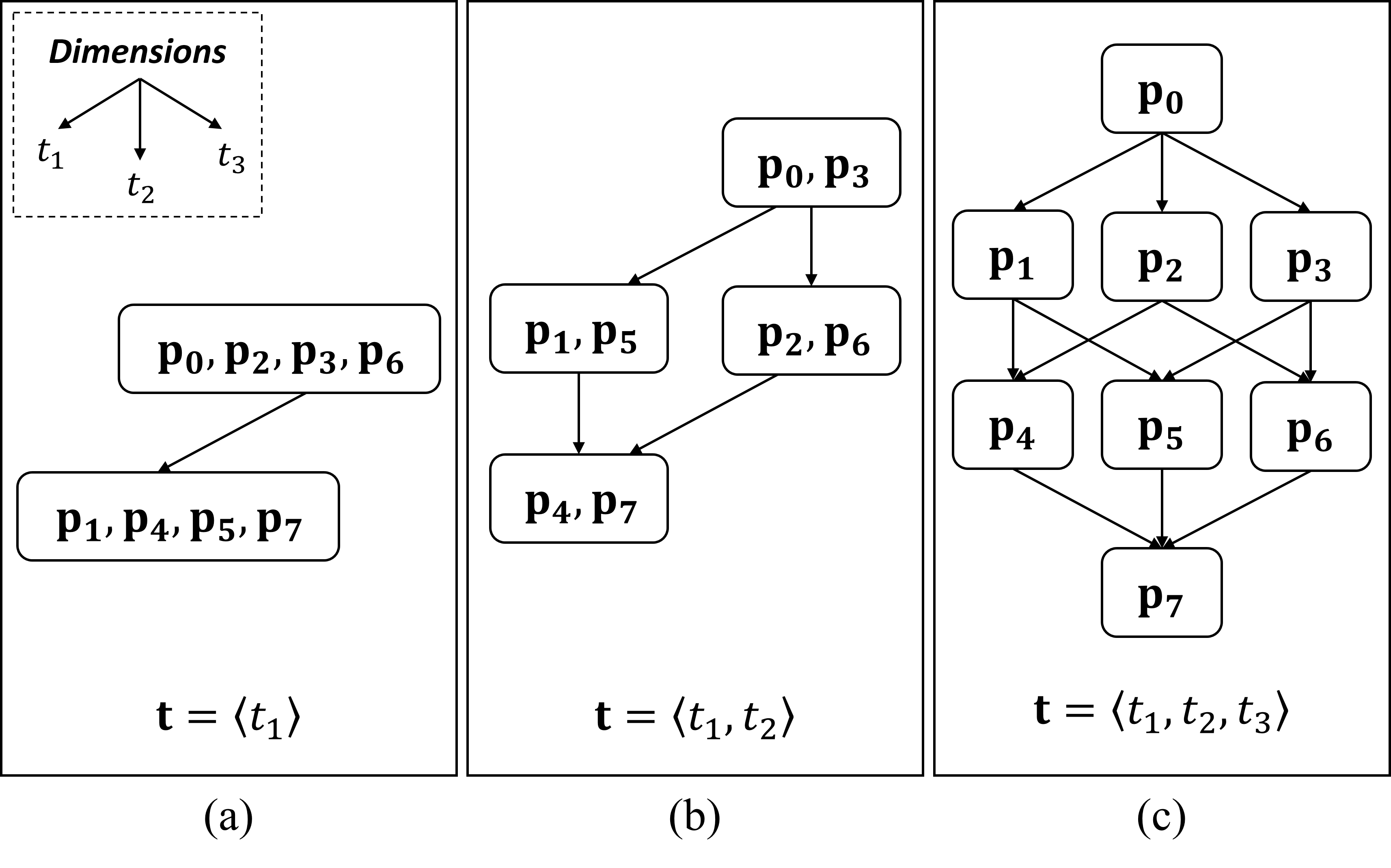}
	\caption{Growth of the PDL from one-dimension to three-dimension}
	\label{fig:growth}
\end{figure}

PDL is worthwhile for considering the relationship between the growth of tests and positions. To put it bluntly, PDL grows by adding tests. Figure \ref{fig:growth} demonstrates the growth of PDL from one-dimension to three-dimension. In Figure \ref{fig:growth} (c), the PDL is the same as Figure \ref{fig:graph}, which represents a three-dimensional program space. Figure \ref{fig:growth} (a) and (b) show the growth of the PDL as tests are individually added. Note that the eight positions $\mathbf{p_0}, \cdots, \mathbf{p_7}$ are expressed in each PDL. First, when the test set $\mathbf{t}$ has only one test $t_1$, $\mathbf{p_1, p_4, p_5,p_7}$ are deviant from $\mathbf{p_0}$ but $\mathbf{p_0, p_2, p_3,p_6}$ are not deviant to each other. After $t_2$ is added, $\mathbf{p_2}$ and $\mathbf{p_6}$ are deviant from $\mathbf{p_0}$, but $\mathbf{p_3}$ is not yet deviant from $\mathbf{p_0}$. Similarly, $\mathbf{p_4}$ and $\mathbf{p_7}$ are deviant from $\mathbf{p_1,p_5}$, and there are four different nodes in Figure \ref{fig:graph} (b). Finally, $t_3$ makes $\mathbf{p_3}$ deviant from $\mathbf{p_0}$, $\mathbf{p_5}$ from $\mathbf{p_1}$, $\mathbf{p_6}$ from $\mathbf{p_2}$, and $\mathbf{p_7}$ from $\mathbf{p_4}$, respectively. 

It is possible to consider testing process as the growth of PDL. For example, in mutation testing, the generation of new tests and execution of live mutants are repeated until the set of tests kills all of the mutants. This means that the generation of new dimensions is repeated until all of the positions of mutants are deviant from the origin. 

Mathematically, PDL is a form of hypercube graph, with $2^n$ nodes, $2^{n-1}n$ edges, and $n$ edges touching each node, where $n$ is the number of dimensions in the corresponding program space. When PDL is applied to mutation-based testing methods, several mathematical properties of hypercube graphs may guide for the elaboration of the mutation-based testing methods.
\section{Applications}\label{sec:applications}

\subsection{Mutation-based Fault Localization}\label{sec:MBFL}
Program faults can be detected by fault detection methods such as mutation testing; once a fault is detected, its location must be analyzed so that the fault can be corrected. This can be very tedious and time consuming, especially in large and complex programs. Among the many contributions to the field of fault localization, Spectrum-Based Fault Localization (SBFL) has received significant attention, owing to its simplicity and effectiveness \cite{xie2013theoretical}. A program spectrum contains information recorded during the execution of a program, such as its executed statements. The spectrum is used to identify suspicious statements that cause a program test to fail. The basic idea is that executed statements that cause the failed test are associated with the failure. For example, if only one executed statement is associated with a failed test, it is obvious that the executed statement caused the test to fail; thus, the location of the fault is that statement. After many tests are executed, the suspiciousness value of each statement is calculated based on the similarity of the program spectrum and the testing results (i.e., pass or fail). All statements are ordered by their suspiciousness value. The higher a statement's suspiciousness value, the higher its probability of being faulty. An ideal fault localization method ranks the faulty statement at the top, to allow programmers or even repair algorithms to correct the fault. There are many suspiciousness (i.e., similarity) metrics, such as Tarantula \cite{jones2005empirical}, Ochiai \cite{abreu2006evaluation}, and Jaccard \cite{chen2002pinpoint}. Recently, Xie et al. \cite{xie2013theoretical} developed a theoretical framework to analyze the efficacy of suspiciousness metrics.

Recently, several researchers developed a new fault localization concept called Mutation-Based Fault Localization (MBFL) \cite{papadakis2013metallaxis,moon2014ask}. Similar to SBFL, MBFL calculates the suspiciousness value of each statement in a program, and ranks the statements in the order of their suspiciousness value. The key feature of MBFL is that the suspiciousness of each statement is calculated according to the suspiciousness of the mutants in the statement. Because many mutants are generated from each statement in general, MBFL has finer granularity than SBFL.

Interestingly, according to the experiments in \cite{papadakis2013metallaxis,moon2014ask}, MBFL has significant advantages over SBFL. However, analyzing and discussing MBFL can be difficult, because of its lack of formal foundations. Thus, a program space is applied to analyze and discuss the foundations of MBFL. Based on the analysis results, there are two fundamental considerations in MBFL: (1) considering a mutant as a potential fix, and (2) considering a mutant as a fault.

\subsubsection{Mutant as a partial fix}
Seokhyeon et al. \cite{moon2014ask} considered two types of mutants: $m_c$, which represents a mutant generated by mutating a correct statement, and $m_f$, which represents a mutant generated by mutating a faulty statement. They observed that failed tests on the original program $p$ of the mutants are more likely to pass on $m_f$ than on $m_c$. On the other hand, passed tests on $p$ are more likely to fail on $m_c$ than on $m_f$. Let $n_{f\to p}(m)$ be the number of tests that failed on $p$ but passed on an arbitrary mutant $m$. Then, the proportion of $n_{f\to p}(m)$ over all failed tests implies the likelihood of $m=m_f$. Similarly, let $n_{p\to f}(m)$ be the number of tests that passed on $p$ but failed on an arbitrary mutant $m$; then, the proportion of $n_{p\to f}(m)$ over all passed tests implies the likelihood of $m=m_c$. Based on these observations, the suspiciousness value of $m$ is calculated by the likelihood of $m=m_f$ minus the likelihood of $m=m_c$.

In a program space, a test $t$ that failed on $p$ but passed on $m$ corresponds to a dimension $t$ such that:
\begin{displaymath}
\begin{split}
&\mathbf{d}_{p_s}^{\langle t \rangle}(p)=1 \wedge \mathbf{d}_{p_s}^{\langle t \rangle}(m)=0 \\
&\Leftrightarrow (\mathbf{d}_{p_s}^{\langle t \rangle}(p) \neq \mathbf{d}_{p_s}^{\langle t \rangle}(m)) \wedge \mathbf{d}_{p_s}^{\langle t \rangle}(m)=0.
\end{split}
\end{displaymath}
Let $\mathbf{t'}$ be a collection of $t$ satisfying $\mathbf{d}_{p_s}^{\langle t \rangle}(p) \neq \mathbf{d}_{p_s}^{\langle t \rangle}(m)$. Then, $n_{f\to p}(m)$ is equal to the number of zeros in $\mathbf{d}_{p_s}^{\mathbf{t'}}(m)$. In the same manner, $n_{p\to f}(m)$ is equal to the number of ones in $\mathbf{d}_{p_s}^{\mathbf{t'}}(m)$. For example, if $\mathbf{d}_{p_s}^{\mathbf{t'}}(m) = \langle 0,1,0,0 \rangle$, then $n_{f\to p}(m)=3$ and $n_{p\to f}(m)=1$. 

This signifies that the suspiciousness value of $m$ increases as the position of $m$ moves towards the origin $p_s$ in the program space. Note that the dimensions of the space are $\mathbf{t'}$, not $\mathbf{t}$. In other words, Seokhyeon et al. \cite{moon2014ask} effectively found the $m$ close to $p_s$ by focusing on the dimensions that caused the test result changes (i.e., $p\to f$ or $f\to p$) for $p$ and $m$.

\subsubsection{Mutant as a fault}
Papadakis and Traon \cite{papadakis2013metallaxis} considered the conduct\footnote{In \cite{papadakis2013metallaxis}, the term \textit{behavior} is used, not \textit{conduct}. Because \textit{behavior} is used as another means in this paper, here we use an alternative term \textit{conduct} to avoid confusion.} of faults (including mutants) with regard to tests. They affirmed that a mutant $m_x$ has the \textit{same conduct} as another mutant $m_y$ if $m_x$ and $m_y$ are killed by the same tests. The key assumption of \cite{papadakis2013metallaxis} is that mutants and faults located on the same program statement will show similar motions. Based on this assumption, for a mutant $m$ and an unlocalized fault $f$, the location of $f$ is given by $m$, whose conduct is similar to the action of $f$. In other words, the suspiciousness value of $m$ is calculated according to the similarity between the conduct of $m$ and the conduct of $f$.

Unfortunately, this assumption is insufficient for calculating the suspiciousness value of $m$ when the conduct of faulty program $p_o$ is not clearly defined. In \cite{papadakis2013metallaxis}, the conduct of $p_o$ is implicitly defined as test results (i.e., pass or fail) instead of kill results. This signifies that the conduct of $p_o$ is based on $p_s$, while the conduct of $m$ is based on $p$. The meaning of the similarity between the conduct of $m$ and the conduct of $p_o$ is ambiguous, because they have different bases. A program space is applied to explore this ambiguousness.

In a program space, the conduct of a mutant $m$ is
\begin{displaymath}
\mathbf{d}_{p_o}^{\mathbf{t}}(m)
\end{displaymath}
for an original program $p_o$, a set of tests (as a test vector) $\mathbf{t}$, and a differentiator $d$. This indicates that the conduct of $m$ represents the position of the point of $m$ relative to the origin $p_o$. On the other hand, based on the implicit definition, the conduct of a faulty program $p$ is
\begin{displaymath}
\mathbf{d}_{p_s}^{\mathbf{t}}(p_o)
\end{displaymath}
for a correct program $p_s$, $\mathbf{t}$, and $d$. This indicates that the conduct of $p_o$ represents the position of $p_o$ relative to the origin $p_s$, not $p_o$. As a result, the suspiciousness value of $m$ is calculated according to the following similarity:
\begin{displaymath}
\begin{split}
&\mathbf{d}_{p_o}^{\mathbf{t}}(m) \sim \mathbf{d}_{p_s}^{\mathbf{t}}(p_o) \\
&{}={} \mathbf{d}_{p_o}^{\mathbf{t}}(m) \sim \mathbf{d}_{p_o}^{\mathbf{t}}(p_s).
\end{split}
\end{displaymath}
This signifies that the suspiciousness value of $m$ represents the proximity between the positions of $m$ and the position of $p_s$ in the space whose origin is $p_o$. In other words, while it is not explained in \cite{papadakis2013metallaxis}, $m$ and $p_s$ are regarded as two faulty programs based on $p_o$.

\subsubsection{Implications}

\begin{figure}
    \centering
    \subfigure[Mutant as a partial fix]{\label{fig:mbfl1}\includegraphics[height=0.15\textheight]{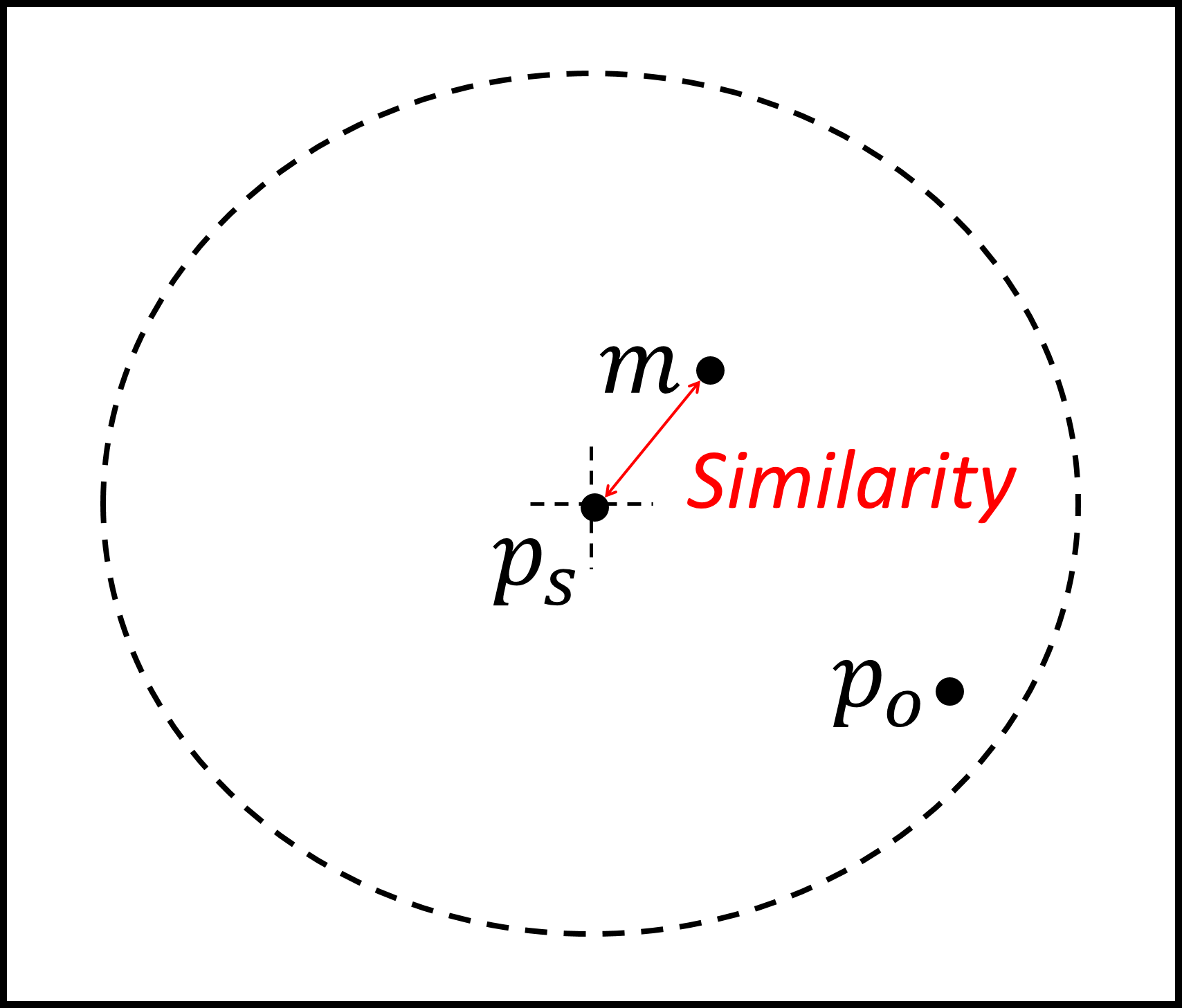}}
    \subfigure[Mutant as a fault]{\label{fig:mbfl2}\includegraphics[height=0.15\textheight]{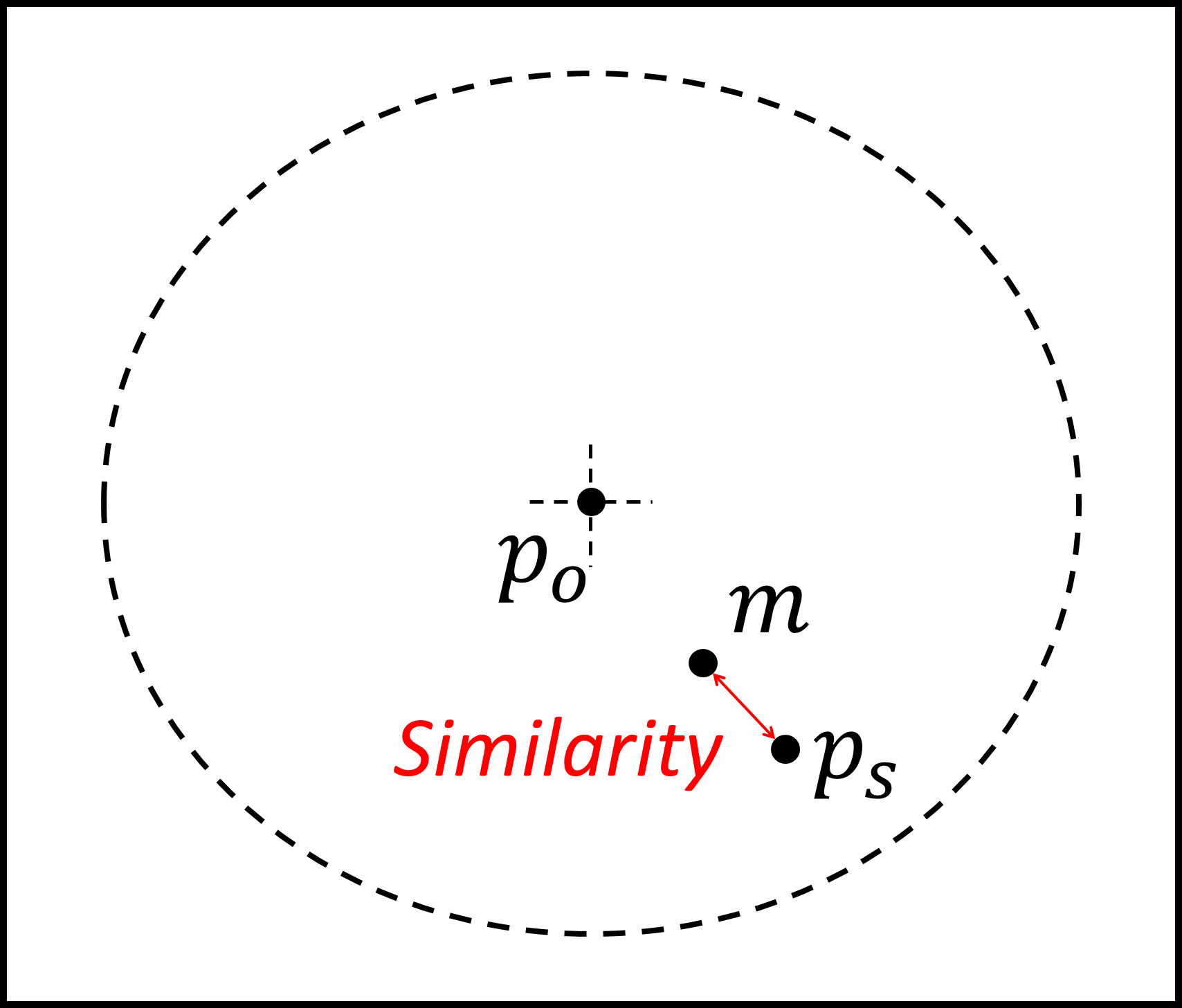}}
    \caption{Two foundations of MBFL represented by program spaces}
    \label{fig:mbfl}
\end{figure}

The conceptual foundations of MBFL are summarized in Figure \ref{fig:mbfl}. Each of the dotted circles represents a program space, and the crossed line at the center implies the origin of each space. Each program space contains the three major points corresponding to $p_o$, $m$, and $p_s$. The arrow represents the method for calculating the suspiciousness value of $m$. For simplicity, MBFL-FIX refers to the MBFL methods that consider a mutant as a partial fix, and MBFL-FLT refers to the MBFL methods that consider a mutant as a fault. Figure \ref{fig:mbfl1} shows the foundation of MBFL-FIX, and Figure \ref{fig:mbfl1} shows the foundation of MBFL-FLT.

It is clearly shown that both MBFL methods focus on the same objective: the similarity (or proximity) between $m$ and $p_s$. This indicates that MBFL is fundamentally the same as program repair, in which the objective is to move toward $p_s$ from $p$ using $m$. This implication matches our intuition, because the objective of fault localization is to find the location that requires the correction. As $m$ moves closer to $p_s$, the location of $m$ in the program is more likely to be the location requiring the correction. There are even cases in which $m$ is the direct correction of $p_o$ in both MBFL methods \cite{papadakis2013metallaxis,moon2014ask}.

Regarding the differences between the two methods, MBFL-FIX utilizes $\mathbf{d}_{p_s}^{\mathbf{t}}(m) \sim \mathbf{0}$, while MBFL-FLT utilizes $\mathbf{d}_{p_o}^{\mathbf{t}}(m) \sim \mathbf{d}_{p_o}^{\mathbf{t}}(p_s)$. As analyzed in (\ref{eq:diff2pos}), $\mathbf{d}_{p_o}^{\mathbf{t}}(m) \sim \mathbf{d}_{p_o}^{\mathbf{t}}(p_s)$ $\centernot\implies$ $\mathbf{d}_{p_s}^{\mathbf{t}}(m) \sim \mathbf{0}$. Specifically, if $p_o, p_s$, and $m$ are different from each other in a test $t$, MBFL-FLT will determine that $m$ is the same as $p_s$ for $t$, while MBFL-FIX will determine that $m$ is different from $p_s$ for $t$. This shows that MBFL-FIX is, at least, better than MBFL-FLT for calculating the similarity between $m$ and $p_s$. We believe additional theoretical and experimental studies will result in improved MBFL methods.

\subsection{Mutant Set Minimization}\label{sec:DMSG}
One long-standing problem that prevents mutation-based testing from becoming practical is the high cost of executing a very large number of mutants against a set of tests \cite{jia2011analysis}. In terms of reducing cost, minimizing mutant sets by removing redundant mutants (with respect to given tests) is a promising strategy. In this subsection, we introduce a mutant set minimization method for a given set of tests, and apply a position deviance lattice (PDL) to provide deeper implications.

Recently, Ammann et al. \cite{ammann2014establishing} established a theoretical foundation for mutant set minimization based on the formal relations of mutants, called dynamic subsumption. If a mutant $m_x$ is killed by at least one test in a set of tests $TS$ and another mutant $m_y$ is always killed whenever $m_x$ is killed, then $m_x$ dynamically subsumes $m_y$ with respect to $TS$. They provided that a mutant set $M_{min}$ is minimal with respect to $TS$ if and only if there does not exist a distinct pair $m_x,m_y\in M_{min}$ such that $m_x$ dynamically subsumes $m_y$. This signifies that if $m_x$ dynamically subsumes $m_y$, then $m_y$ is redundant to $m_x$ with respect to $TS$. For example, consider four mutants $m_1, m_2, m_3, and m_4$, and a set of tests $\mathbf{t}=\langle t_1,t_2,t_3 \rangle$. Table \ref{table:mutants} shows which mutants each test kills. Specifically, the $(i,j)$ element of the table is the value of $d(t_i,p_o,m_j)$: if $t_i$ kills $m_j$ then 1, otherwise 0.

\begin{table}
\caption{Example: Mutant Kill Information}
\label{table:mutants}
    \begin{center}
        \begin{tabular}{|c|cccc|}
        \hline
         & $m_1$ & $m_2$ & $m_3$ & $m_4$ \\
        \hline
		$t_1$ & 1&	0&	1&	1 \\
		$t_2$ & 0&	1&	0&	1 \\
		$t_3$ & 0&	1&	1&	1 \\
        \hline
        \end{tabular}
    \end{center}
\end{table}

According to the definition of dynamic subsumption, $m_1$ dynamically subsumes $m_3$ and $m_4$, $m_2$ dynamically subsumes $m_4$, and $m_3$ dynamically subsumes $m_4$. By removing all dynamically subsumed mutants, $M_{min} = \{m_1, m_2\}$ becomes the minimal set of mutants with respect to the test set $TS=\{t_1,t_2,t_3\}$. This provides a solid theoretical foundation for mutant set minimization. We used our theoretical framework to interpret this foundational study, and found more implications to elaborate the mutant set minimization.

\begin{figure}
	\centering
	\includegraphics[width=0.9\linewidth]{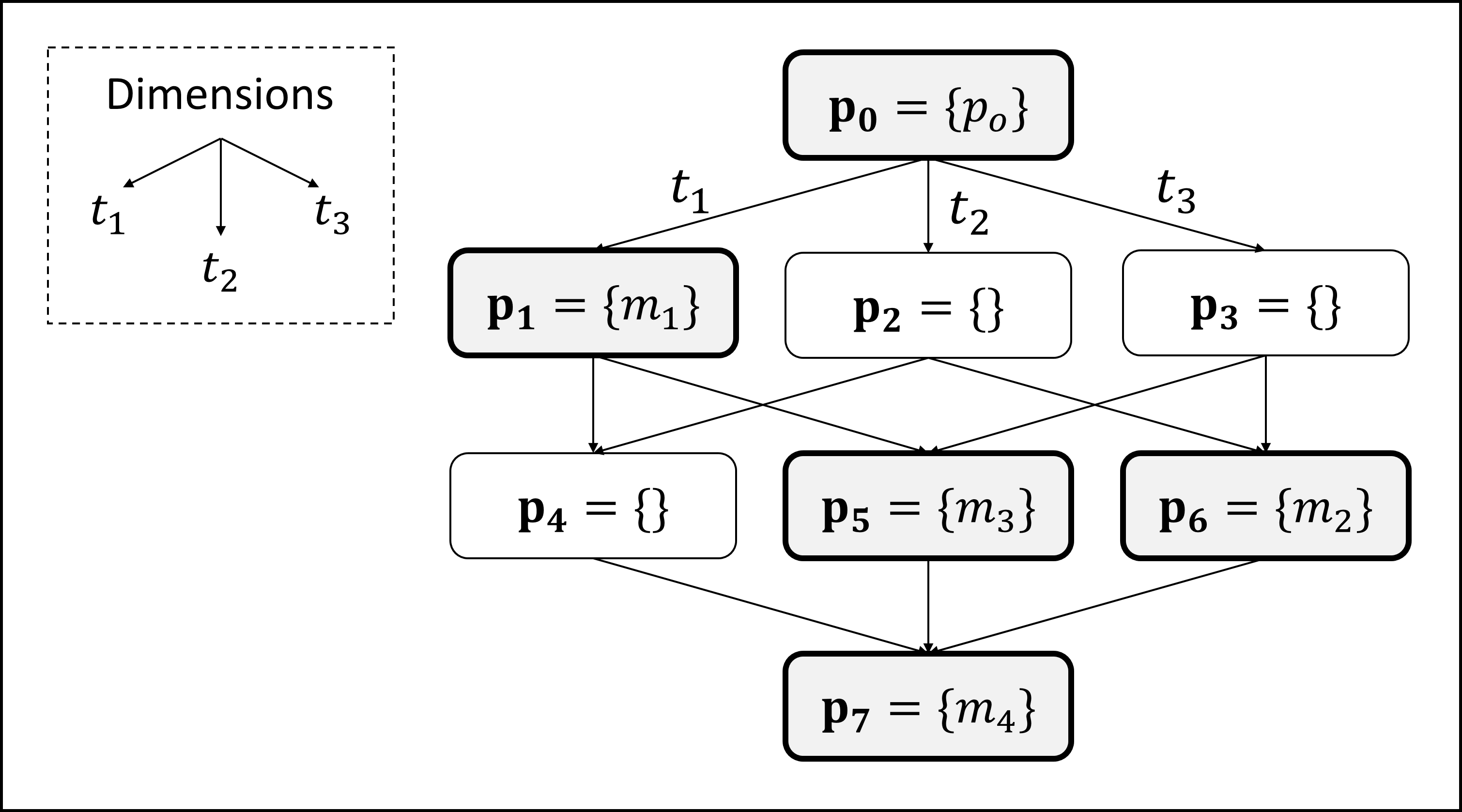}
	\caption{PDL with mutants for finding a minimal set of mutants}
	\label{fig:graph-mutants}
\end{figure}

Let us create a PDL based on Table \ref{table:mutants}. As we previously discussed in Section \ref{sec:position}, $\mathbf{d}_{p_o}^{\mathbf{t}}(m) = \mathbf{m}$ implies which $t\in \mathbf{t}$ kills $m$. From Table \ref{table:mutants}, $\mathbf{m_1} = \langle 1,0,0 \rangle$, $\mathbf{m_2} = \langle 0,1,1 \rangle$, $\mathbf{m_3} = \langle 1,0,1 \rangle$, and $\mathbf{m_4} = \langle 1,1,1 \rangle$. This provides the PDL, as shown in Figure \ref{fig:graph-mutants}. The gray box refers to the position that contains a mutant. In the PDL, it is easy to see that $\mathbf{m_1} \rightarrow \mathbf{m_3}$, $\mathbf{m_1} \rightarrow \mathbf{m_4}$, $\mathbf{m_2} \rightarrow \mathbf{m_4}$, and $\mathbf{m_3} \rightarrow \mathbf{m_4}$. Note that the deviance relations between positions precisely correspond to the dynamic subsumption relations between mutants. Formally, the following is true for all $p_o\in P$, $m_x,m_y\in M_{min}$ generated from $p_o$, $\mathbf{t}$, and $d$:
\begin{align}
\label{eq:1}	\mathbf{p_o} &\rightarrow \mathbf{m_x} \rightarrow \mathbf{m_y} \text{ in a program space of }(\mathbf{t}, p_o, d)\\
\label{eq:2}			   \Leftrightarrow{}&\exists \mathbf{t_d}\subseteq \mathbf{t} \text{ } (\mathbf{d}_{p_o}^{\mathbf{t}}(m_x) \xrightarrow{\mathbf{t_d}} \mathbf{d}_{p_o}^{\mathbf{t}}(m_y))\\
\label{eq:3}										 &{}\wedge{} (\mathbf{d}_{p_o}^{\mathbf{t}}(m_x) \neq \mathbf{0}) \\
\label{eq:4}			   \Leftrightarrow{}&\forall t\in \mathbf{t}\text{ } ( (d(t,p_o,m_x)=1) \implies (d(t,p_o,m_y)=1) ) \\
\label{eq:5}										 &{}\wedge{} \exists t\in \mathbf{t} \text{ } d(t,p_o,m_x)=1 \\
\label{eq:6}			   \Leftrightarrow{}& m_x \text{ dynamically subsumes } m_y \text{ with respect to } \mathbf{t}
\end{align}

Briefly, (\ref{eq:2}) signifies $\mathbf{m_x} \xrightarrow{\mathbf{t_d}} \mathbf{m_y}$ for some $\mathbf{t_d} \neq \emptyset$, and (\ref{eq:3}) indicates that $m_x$ is killed by at least one test because its position is deviant from $\mathbf{p_o}$. (\ref{eq:4}) corresponds to (\ref{eq:2}), and (\ref{eq:5}) corresponds to (\ref{eq:4}). Consequently, (\ref{eq:1}) is equivalent to (\ref{eq:6}). This signifies that the deviance relation on positions in a certain program space is the same as the dynamic subsumption relation on mutants. In other words, deviance relations are a more general and comprehensive concept for representing the behavioral differences of programs, while dynamic subsumption relations are a more specific concept for representing the behavioral differences between mutants and an original program. The comprehensiveness of deviance relations is also shown in Figure \ref{fig:graph-mutants}. If we remove all nodes that do not have a mutant from the PDL, the remaining graph will precisely represent the dynamic mutant subsumption graph (DMSG) introduced by Kurtz et al. \cite{kurtz2014mutant}. The DMSG makes it easy to identify $M_{min}$, which is simply the set of root nodes; of course, the PDL provides the same benefit.

The PDL provides more implications than the DMSG. In Figure \ref{fig:graph-mutants}, the maximum number of positions that are not deviant from the other is three. In other words, for an arbitrary four (or more) positions, at least one position is deviant from another position. This indicates that the theoretical maximum size of $M_{min}$ is three, with respect to any test set $|TS|=3$. In general, the maximum size of $M_{min}$ is
\begin{displaymath}
\max(|M_{min}|) = \binom{n}{\floor{n/2}}
\end{displaymath}
where $n$ is the size of a test set. In other words, for an arbitrary test set $TS$ and an arbitrary mutants set $M$, the theoretical maximum size of the minimal mutant set $M_{min}$ with respect to $TS$ is $\binom{n}{\floor{n/2}}$. Further, $\max(|M_{min}|)$ is given without executing all $m\in M$ against $TS$. For example, for $|TS|=5$, $\max(|M_{min}|)$ is $\binom{5}{2} = 10$. This signifies that, with respect to the given $TS$, any mutant set $M$ will be reduced to contain at most 10 mutants. This improves the theoretical foundation for mutant set minimization by highlighting the relationship between the size of $M_{min}$ and $TS$. Yet, the relationship between $TS$ and $M$ should be investigated further to determine, for example, the meaning of the ratio of $|M_{min}|$ to $\max(|M_{min}|)$, or the practical maximum of $|M_{min}|$ with respect to $TS$, based on the theoretical framework.
\section{Conclusion}\label{sec:conclusion}

In this paper, we considered a theoretical framework for better understanding of mutation-based testing methods. In particular, we defined a test differentiator to shift the paradigm of testing from the correctness of a program to the difference between programs. A test differentiator clearly and concisely represents the behavioral differences between programs in a test. With regards to a test set, we defined a d-vector that represents the behavioral differences between two programs in vector form. 

Using the fact that a vector can be regarded as representing a point in a multidimensional space, we define the space of programs corresponding to d-vectors. In the program space, the position of a program relative to the origin in each dimension indicates the behavioral difference between the program and the origin for the test corresponding to the dimension. The relationship between different positions and behaviors is clearly addressed. We then continued to define the derivation relation on positions for representing how tests influence positions in a testing process. The position derivation lattice (PDL) is defined for providing visual aids for positions and their derivation relations. 

We then revisited the existing mutation-based fault localization methods and the mutant set minimization method, demonstrating the applicability of our theoretical framework. For mutation-based fault localization methods, we found that the common foundation is the proximity between mutants and the correct program in the program space, while the method of calculating the proximity is different. We also found that one method is, at least, theoretically better than the other method. Furthermore, we showed that our theoretical framework is sufficiently general to include all theoretical foundations for mutant set minimization. We also improved the mutant set minimization theory by providing the theoretical maximum size of a minimal mutant set. Given our results, we demonstrated that our theoretical framework may serve as a solid foundation for discussions in both empirical and theoretical studies on mutation-based testing.




\bibliographystyle{IEEEtran}
\bibliography{IEEEabrv,MutationFoundation}

\begin{thebibliography}{10}
\providecommand{\url}[1]{#1}
\csname url@samestyle\endcsname
\providecommand{\newblock}{\relax}
\providecommand{\bibinfo}[2]{#2}
\providecommand{\BIBentrySTDinterwordspacing}{\spaceskip=0pt\relax}
\providecommand{\BIBentryALTinterwordstretchfactor}{4}
\providecommand{\BIBentryALTinterwordspacing}{\spaceskip=\fontdimen2\font plus
\BIBentryALTinterwordstretchfactor\fontdimen3\font minus
  \fontdimen4\font\relax}
\providecommand{\BIBforeignlanguage}[2]{{%
\expandafter\ifx\csname l@#1\endcsname\relax
\typeout{** WARNING: IEEEtran.bst: No hyphenation pattern has been}%
\typeout{** loaded for the language `#1'. Using the pattern for}%
\typeout{** the default language instead.}%
\else
\language=\csname l@#1\endcsname
\fi
#2}}
\providecommand{\BIBdecl}{\relax}
\BIBdecl

\bibitem{demillo1978hints}
R.~A. DeMillo, R.~J. Lipton, and F.~G. Sayward, ``Hints on test data selection:
  Help for the practicing programmer,'' \emph{Computer}, vol.~11, no.~4, pp.
  34--41, 1978.

\bibitem{andrews2006using}
J.~H. Andrews, L.~C. Briand, Y.~Labiche, and A.~S. Namin, ``Using mutation
  analysis for assessing and comparing testing coverage criteria,''
  \emph{Software Engineering, IEEE Transactions on}, vol.~32, no.~8, pp.
  608--624, 2006.

\bibitem{li2009experimental}
N.~Li, U.~Praphamontripong, and J.~Offutt, ``An experimental comparison of four
  unit test criteria: Mutation, edge-pair, all-uses and prime path coverage,''
  in \emph{Software Testing, Verification and Validation Workshops, 2009.
  ICSTW'09. International Conference on}.\hskip 1em plus 0.5em minus
  0.4em\relax IEEE, 2009, pp. 220--229.

\bibitem{just2014mutants}
R.~Just, D.~Jalali, L.~Inozemtseva, M.~D. Ernst, R.~Holmes, and G.~Fraser,
  ``Are mutants a valid substitute for real faults in software testing?'' in
  \emph{Proceedings of the 22nd ACM SIGSOFT International Symposium on
  Foundations of Software Engineering}.\hskip 1em plus 0.5em minus 0.4em\relax
  ACM, 2014, pp. 654--665.

\bibitem{schulte2014software}
E.~Schulte, Z.~P. Fry, E.~Fast, W.~Weimer, and S.~Forrest, ``Software
  mutational robustness,'' \emph{Genetic Programming and Evolvable Machines},
  vol.~15, no.~3, pp. 281--312, 2014.

\bibitem{papadakis2013metallaxis}
M.~Papadakis and Y.~Le~Traon, ``Metallaxis-fl: mutation-based fault
  localization,'' \emph{Software Testing, Verification and Reliability}, 2013.

\bibitem{moon2014ask}
S.~Moon, Y.~Kim, M.~Kim, and S.~Yoo, ``Ask the mutants: Mutating faulty
  programs for fault localization,'' in \emph{Software Testing, Verification
  and Validation (ICST), 2014 IEEE Seventh International Conference on}.\hskip
  1em plus 0.5em minus 0.4em\relax IEEE, 2014, pp. 153--162.

\bibitem{debroy2010using}
V.~Debroy and W.~E. Wong, ``Using mutation to automatically suggest fixes for
  faulty programs,'' in \emph{Software Testing, Verification and Validation
  (ICST), 2010 Third International Conference on}.\hskip 1em plus 0.5em minus
  0.4em\relax IEEE, 2010, pp. 65--74.

\bibitem{le2012genprog}
C.~Le~Goues, T.~Nguyen, S.~Forrest, and W.~Weimer, ``Genprog: A generic method
  for automatic software repair,'' \emph{Software Engineering, IEEE
  Transactions on}, vol.~38, no.~1, pp. 54--72, 2012.

\bibitem{debroy2014combining}
V.~Debroy and W.~E. Wong, ``Combining mutation and fault localization for
  automated program debugging,'' \emph{Journal of Systems and Software},
  vol.~90, pp. 45--60, 2014.

\bibitem{gourlay1983mathematical}
J.~S. Gourlay, ``A mathematical framework for the investigation of testing,''
  \emph{Software Engineering, IEEE Transactions on}, no.~6, pp. 686--709, 1983.

\bibitem{goodenough1975toward}
J.~B. Goodenough and S.~L. Gerhart, ``Toward a theory of test data selection,''
  \emph{Software Engineering, IEEE Transactions on}, no.~2, pp. 156--173, 1975.

\bibitem{howden1976reliability}
W.~E. Howden, ``Reliability of the path analysis testing strategy,''
  \emph{Software Engineering, IEEE Transactions on}, no.~3, pp. 208--215, 1976.

\bibitem{weyuker1980theories}
E.~J. Weyuker and T.~J. Ostrand, ``Theories of program testing and the the
  application of revealing subdomains.'' \emph{Software Engineering, IEEE
  Transactions on}, vol.~6, no.~3, pp. 236--246, 1980.

\bibitem{staats2011programs}
M.~Staats, M.~W. Whalen, and M.~P.~E. Heimdahl, ``Programs, tests, and oracles:
  the foundations of testing revisited,'' in \emph{Software Engineering (ICSE),
  2011 33rd International Conference on}.\hskip 1em plus 0.5em minus
  0.4em\relax IEEE, 2011, pp. 391--400.

\bibitem{staats2012understanding}
M.~Staats, S.~Hong, M.~Kim, and G.~Rothermel, ``Understanding user
  understanding: determining correctness of generated program invariants,'' in
  \emph{Proceedings of the 2012 International Symposium on Software Testing and
  Analysis}.\hskip 1em plus 0.5em minus 0.4em\relax ACM, 2012, pp. 188--198.

\bibitem{polikarpova2013good}
N.~Polikarpova, C.~A. Furia, Y.~Pei, Y.~Wei, and B.~Meyer, ``What good are
  strong specifications?'' in \emph{Proceedings of the 2013 International
  Conference on Software Engineering}.\hskip 1em plus 0.5em minus 0.4em\relax
  IEEE Press, 2013, pp. 262--271.

\bibitem{fraser2014achieving}
G.~Fraser and A.~Arcuri, ``Achieving scalable mutation-based generation of
  whole test suites,'' \emph{Empirical Software Engineering}, vol.~20, no.~3,
  pp. 783--812, 2014.

\bibitem{barr2015oracle}
E.~T. Barr, M.~Harman, P.~McMinn, M.~Shahbaz, and S.~Yoo, ``The oracle problem
  in software testing: A survey,'' 2015.

\bibitem{offutt1992investigations}
A.~J. Offutt, ``Investigations of the software testing coupling effect,''
  \emph{ACM Transactions on Software Engineering and Methodology (TOSEM)},
  vol.~1, no.~1, pp. 5--20, 1992.

\bibitem{budd1980theoretical}
T.~A. Budd, R.~A. DeMillo, R.~J. Lipton, and F.~G. Sayward, ``Theoretical and
  empirical studies on using program mutation to test the functional
  correctness of programs,'' in \emph{Proceedings of the 7th ACM SIGPLAN-SIGACT
  symposium on Principles of programming languages}.\hskip 1em plus 0.5em minus
  0.4em\relax ACM, 1980, pp. 220--233.

\bibitem{budd1982two}
T.~A. Budd and D.~Angluin, ``Two notions of correctness and their relation to
  testing,'' \emph{Acta Informatica}, vol.~18, no.~1, pp. 31--45, 1982.

\bibitem{morell1990theory}
L.~J. Morell, ``A theory of fault-based testing,'' \emph{Software Engineering,
  IEEE Transactions on}, vol.~16, no.~8, pp. 844--857, 1990.

\bibitem{harman2011strong}
M.~Harman, Y.~Jia, and W.~B. Langdon, ``Strong higher order mutation-based test
  data generation,'' in \emph{Proceedings of the 19th ACM SIGSOFT symposium and
  the 13th European conference on Foundations of software engineering}.\hskip
  1em plus 0.5em minus 0.4em\relax ACM, 2011, pp. 212--222.

\bibitem{offutt2011mutation}
J.~Offutt, ``A mutation carol: Past, present and future,'' \emph{Information
  and Software Technology}, vol.~53, no.~10, pp. 1098--1107, 2011.

\bibitem{woodward1988weak}
M.~Woodward and K.~Halewood, ``From weak to strong, dead or alive? an analysis
  of some mutation testing issues,'' in \emph{Software Testing, Verification,
  and Analysis, 1988., Proceedings of the Second Workshop on}.\hskip 1em plus
  0.5em minus 0.4em\relax IEEE, 1988, pp. 152--158.

\bibitem{ammann2014establishing}
P.~Ammann, M.~E. Delamaro, and J.~Offutt, ``Establishing theoretical minimal
  sets of mutants,'' in \emph{Software Testing, Verification and Validation
  (ICST), 2014 IEEE Seventh International Conference on}.\hskip 1em plus 0.5em
  minus 0.4em\relax IEEE, 2014, pp. 21--30.

\bibitem{xie2013theoretical}
X.~Xie, T.~Y. Chen, F.-C. Kuo, and B.~Xu, ``A theoretical analysis of the risk
  evaluation formulas for spectrum-based fault localization,'' \emph{ACM
  Transactions on Software Engineering and Methodology (TOSEM)}, vol.~22,
  no.~4, p.~31, 2013.

\bibitem{jones2005empirical}
J.~A. Jones and M.~J. Harrold, ``Empirical evaluation of the tarantula
  automatic fault-localization technique,'' in \emph{Proceedings of the 20th
  IEEE/ACM international Conference on Automated software engineering}.\hskip
  1em plus 0.5em minus 0.4em\relax ACM, 2005, pp. 273--282.

\bibitem{abreu2006evaluation}
R.~Abreu, P.~Zoeteweij, and A.~J. Van~Gemund, ``An evaluation of similarity
  coefficients for software fault localization,'' in \emph{Dependable
  Computing, 2006. PRDC'06. 12th Pacific Rim International Symposium on}.\hskip
  1em plus 0.5em minus 0.4em\relax IEEE, 2006, pp. 39--46.

\bibitem{chen2002pinpoint}
M.~Y. Chen, E.~Kiciman, E.~Fratkin, A.~Fox, and E.~Brewer, ``Pinpoint: Problem
  determination in large, dynamic internet services,'' in \emph{Dependable
  Systems and Networks, 2002. DSN 2002. Proceedings. International Conference
  on}.\hskip 1em plus 0.5em minus 0.4em\relax IEEE, 2002, pp. 595--604.

\bibitem{jia2011analysis}
Y.~Jia and M.~Harman, ``An analysis and survey of the development of mutation
  testing,'' \emph{Software Engineering, IEEE Transactions on}, vol.~37, no.~5,
  pp. 649--678, 2011.

\bibitem{kurtz2014mutant}
B.~Kurtz, P.~Ammann, M.~E. Delamaro, J.~Offutt, and L.~Deng, ``Mutant
  subsumption graphs,'' in \emph{Software Testing, Verification and Validation
  Workshops (ICSTW), 2014 IEEE Seventh International Conference on}.\hskip 1em
  plus 0.5em minus 0.4em\relax IEEE, 2014, pp. 176--185.

\end{thebibliography}

\end{document}